\documentclass[global,twocolumn]{svjour}
\usepackage{graphics}
\usepackage{amsmath}
\journalname{Appl. Phys. B (2008) 92: 615-621}
\begin{document}
\title{Velocity-selective sublevel resonance of atoms 
with an array of current-carrying wires
}
\author{Atsushi Hatakeyama
}                     
%
%
\institute{Department of Applied Physics, Tokyo University of Agriculture and Technology, Koganei, Tokyo 184-8588, Japan \\ Fax: +81-42-388-7554, \email{hatakeya@cc.tuat.ac.jp}, Tel: +81-42-388-7554}
%
%
\maketitle
\begin{abstract}
Resonance transitions between the Zeeman sublevels of optically-polarized Rb atoms traveling through a spatially periodic magnetic field are investigated in a radio-frequency (rf) range of sub-MHz. 
The atomic motion induces the resonance when the Zeeman splitting is equal to the frequency at which the moving atoms feel the magnetic field oscillating. 
Additional temporal oscillation of the spatially periodic field splits a motion-induced resonance peak into two by an amount of this oscillation frequency.
At higher oscillation frequencies, it is more suitable 
to consider that the resonance is mainly driven by the temporal field oscillation, with its velocity-dependence or Doppler shift caused by the
atomic motion through the periodic field. 
A theoretical description of motion-induced resonance is also given, with emphasis on the translational energy change associated with the internal transition.
\end{abstract}

\section{Introduction}
\label{intro}
The study of the interactions of particles with periodic 
structures has a long history and has 
led to many applications in various fields.
The most widely-known are perhaps diffractions by crystals
for electrons and neutrons, 
for both of which the Nobel Prizes were awarded in 1937 and 1994, respectively~\cite{Nobel37,Nobel94}.
Diffraction of atomic or molecular beams by crystals is also an important technique in surface science~\cite{Far98}.
Besides crystals, artificial periodic structures
like gratings have been used as well
for diffraction of a variety of particles from atoms~\cite{Kei88} to relatively large molecules~\cite{Arn99} and clusters~\cite{Sch94}.
Not only for such de Broglie wave diffraction,
but periodic structures can be used also for changing the classical trajectories of particles; an example is a periodically magnetized surface
used as mirrors for slow paramagnetic atoms~\cite{Opa92,Roa95,Sid96,Lau99,Hin99}.
Another application of the interactions of particles with periodic structures is the generation of coherent radiation
by injecting charged particle beams
into crystals~\cite{NIMB05}.
Various coherent phenomena such as parametric X-ray radiation~\cite{Fre95} 
have been studied extensively for realizing
high energy coherent radiation sources. 
Artificial periodic structures have also been used for the same purpose, and
wigglers or undulators are
successfully implemented
for synchrotron light sources and free electron lasers~\cite{Fel05}.

In many examples mentioned above, the internal states of particles are not usually changed during the interactions.
However, the
interactions with periodic structures can also induce the internal transitions
of incident particles resonantly.
The most extensively studied is a phenomenon called ``resonant coherent excitation''~\cite{Dat78}, 
where periodic perturbation experienced by an ion traversing
inside a crystal induces an internal transition of the ion
when the perturbation frequency
corresponds to the energy difference of the relevant states. 
The resonance of this kind has attracted much attention~\cite{Kra96,Gar04}
since the first
proposal by Okorokov~\cite{Oko65}.
Recent progress achieved by using high energy
beams of highly charged ions has shown the possibilities of
application to high resolution spectroscopy and coherent 
control of
these ions in the X-ray region of keV or $10^{18}$~Hz~\cite{Kon06,Nak08},
the order of which is determined by the ion velocity  ($\sim 10^8$~m/s) and the lattice constant ($\sim 10^{-10}$~m).
The extension of the study on this kind of resonance, however, has been very limited~\cite{Had69}. 

We have recently reported on the resonance that has the same
basic principles as resonant coherent excitation but occurs in a quite
different energy range of neV or $10^5$~Hz~\cite{Hat05}:
magnetic resonance between the Zeeman sublevels of 
Rb atoms having a typical 
velocity of $v=150$~m/s in a
periodic magnetic field (period: $a=1$~mm). 
The resonance, called motion-induced resonance in that report,
was observed in a thin cell containing a Rb vapor.
The Rb atoms with a Doppler-selected velocity
were polarized and their magnetic resonance in the
ground state was 
detected with a circularly polarized laser beam.
The periodic magnetic
field was applied with a pair of arrays of parallel current-carrying
wires sandwiching the cell. 
In this paper we extend our study to the resonance that is induced
by the temporal oscillation of the periodic field together with
the motion of the atoms through the field.
This type of resonance can be viewed as ``velocity-selective'' or ``Doppler-shifted"
rf resonance.
The physical processes of the resonance are easily understood in terms of the energy conservation of the system including the atomic
internal and translational states and the rf photons of the oscillating field.
The resonance we have demonstrated has several unique features that the ordinary rf resonance does not have:
a strong atom velocity dependence, a localized periodic
field near the field source, and
a large atomic momentum change originating from a small period of the field.
These features become more prominent by making the field period
smaller, which can be accomplished  by 
such advancing technologies 
as surface nanofabrication and high-density
magnetic recording.
Together with rapidly
progressing atom control and manipulation
techniques near surfaces, e.g., atom chips~\cite{Fol00} and quantum reflection~\cite{Shi01}, our new approach of using periodic
structures to induce resonance transitions of atoms
will find unique applications to their spectroscopy and control,
especially near surfaces.

In this paper we first present a theoretical formalism to describe resonance induced by atomic motion in a spatially periodic field. 
Following a brief description of the experimental setup, we
then show experimental results for both static and temporally
oscillating periodic fields. Observed resonance spectra
are discussed, particularly from the viewpoint of the energy conservation for
the atomic internal and translational degrees of freedom and rf photons.
Other formulations to describe motion-induced resonance are
also mentioned in the appendix. 

\section{Formulation of resonance transitions of atoms moving in a periodic field}
\subsection{Simple derivation of the resonant condition by classically treating atomic motion}
\label{classical}
We consider the simplest case that an atom with two internal levels, the ground state $|1\rangle$ and the excited
state $|2 \rangle$, 
moves through a sinusoidal field with a period $a$ in a one dimensional space.
The energy splitting between $|1\rangle$ and $|2 \rangle$ is $h\nu_{21}=\hbar\omega_{21} ( > 0)$: $h$ is Planck's constant and $\hbar=h/(2\pi)$.
The field may be expressed as
\begin{equation}
F= A\cos(qx),
\end{equation}
where $q=2\pi/a (>0)$.
If one treats the atomic motion classically and determines the position of
the atom at time t to be $x=vt$, where
$v$ is the atom velocity (not necessary positive in general), the field at the atom position is
\begin{equation}
F=A\cos(qvt).
\end{equation}
Thus the atom experiences a field oscillating with a frequency of $q|v|/(2\pi) = |v|/a$.
When this frequency matches with the transition frequency of the states $|1\rangle$
and $|2\rangle$, namely, 
\begin{equation}
\nu_{21} = \frac{|v|}{a}, 
\label{RC}
\end{equation}
a resonance transition occurs as long as the field couples the two states.

If the amplitude $A$ also oscillates as $A=A'\cos(2\pi f_m t)$, the field the atom experiences is
\begin{align}
		F &= A'\cos(2\pi f_m t) \cos(qvt) \notag \\
		       &= \frac{1}{2}A' \left\{\cos \left[2\pi\left(\frac{v}{a}+f_m\right)t\right] + \cos \left[2\pi\left(\frac{v}{a}-f_m\right)t\right]\right\}.
\label{RCfmC}		       
\end{align}
Hence the resonance condition is
\begin{equation}
\nu_{21} = \left|\frac{v}{a} \pm f_m\right|.
\end{equation}

\subsection{Quantum mechanical treatment}
\label{quantum}
The resonance condition and dynamics can be deduced more naturally and rigorously from a
quantum mechanical description including the atomic translational
motion as well as the atomic internal states.
This treatment is especially essential when the de Broglie wavelength is comparable with the field period.
The formalism described here follows a theory of atomic Bragg
scattering from a standing wave light beam\cite{Mar92}.
We set the Hamiltonian of the system to be
\begin{equation}
H = H_0 + V.
\end{equation}
$H_0$ is the Hamiltonian describing the internal and translational states of
the atom:
\begin{equation}
H_0 = \hbar \omega_{21}|2\rangle\langle2| +  \frac{\hat{p}^2}{2m},
\end{equation}
where $\hat{p}$ and $m$ are the momentum operator and the mass
of the atom, respectively.
$V$ represents the interaction of the dipole moment $\hat{d}$ of the atom with the periodic field $F$:
\begin{equation}
V = -\hat{d}F = \hbar\Omega(|2\rangle\langle1|+|1\rangle\langle2|)\cos (qx).
\end{equation}
$\Omega$ is assumed to be real and positive.
We represent an eigenstate of the Hamiltonian $H_0$ as $|j, k\rangle$, where
$j(=1, 2)$ denotes the internal state of the atom, and $\hbar k$ is the translational momentum of the atom.
The matrix elements of the interaction Hamiltonian $V$ in the $|j, k\rangle$ basis are
\begin{align}
		\langle j', k'|V|j, k\rangle &= \frac{\hbar\Omega}{2} \qquad (\mathrm{for}\  j' \not= j, k' = k \pm q), \notag \\
		       &= 0 \qquad (\mathrm{otherwise}).
\end{align}
We now assume the initial state of the atom at the time $t=0$ is $|1, k_0\rangle$. 
The wave function after evolving under the Hamiltonian $H$ may be written as, using the integer $n$, 
\begin{align}
|\Psi(t)\rangle &= \exp\left(-i\frac{\hbar k_0^2}{2m}t\right) \notag \\
&\times \sum_{n} \left(C_{1,n}(t)  |1, k_0+nq\rangle + C_{2,n}(t)|2, k_0+nq\rangle \right).
\end{align}
Note that the probability amplitudes are all zeros at $t=0$ except $C_{1,0} (t=0) = 1$, and
\begin{gather}
C_{1,n} = 0 \qquad (\mathrm{for}\ n: \mathrm{odd}), \notag \\
C_{2,n} = 0 \qquad (\mathrm{for}\ n: \mathrm{even}). 
\end{gather}
Inserting this wave function into the Schr\"{o}dinger equation, we obtain
\begin{align}
\frac{\partial C_{1,n}}{\partial t} 
= &-i\left(\frac{\hbar k_0 q}{m}n + \frac{\hbar q^2}{2m}n^2 \right) C_{1,n} \notag \\
&-i \frac{\Omega}{2} \left( C_{2,n+1} + C_{2,n-1} \right), \\
\frac{\partial C_{2,n}}{\partial t} 
=&-i\left(\omega_{21} +  \frac{\hbar k_0 q}{m}n + \frac{\hbar q^2}{2m}n^2 \right) C_{2,n} \notag \\
&-i\frac{\Omega}{2} \left( C_{1,n+1} + C_{1,n-1} \right).
\end{align}
We here explicitly write several differential equations for the amplitudes with small $|n|$.
\begin{align}
\frac{\partial C_{1,-2}}{\partial t} 
=&i\left(\frac{2\hbar k_0 q}{m} -\frac{2\hbar q^2}{m}  \right)C_{1,-2}  \notag \\
&-i \frac{\Omega}{2} \left(C_{2,-1} + C_{2,-3} \right), \\
\frac{\partial C_{2,-1}}{\partial t} 
= &-i\left(\omega_{21} - \frac{\hbar k_0 q}{m} + \frac{\hbar q^2}{2m}  \right)C_{2,-1}  \notag \\
&-i \frac{\Omega}{2} \left(C_{1,0} + C_{1,-2} \right), \label{C2-1} \\
\frac{\partial C_{1,0}}{\partial t} 
= &-i\frac{\Omega}{2} \left(C_{2,1} + C_{2,-1} \right),\\ 
\frac{\partial C_{2,1}}{\partial t} 
= &-i\left(\omega_{21} + \frac{\hbar k_0 q}{m} + \frac{\hbar q^2}{2m} \right)C_{2,1} \notag \\
&-i \frac{\Omega}{2} \left(C_{1,2} + C_{1,0} \right). 
\end{align}
We now consider the case that $\omega_{21} - \hbar k_0 q/m + \hbar q^2/(2m)=0$, namely, the coefficient of $C_{2,-1}$ in the right hand side of Eq.~(\ref{C2-1}) is zero.
The above coupling equations are then reduced to
\begin{align}
&\frac{\partial C_{1,-2}}{\partial t} 
= i\left(2\omega_{21} - \frac{\hbar q^2}{m} \right)C_{1,-2} -i \frac{\Omega}{2} \left(C_{2,-1} + C_{2,-3} \right), \\
&\frac{\partial C_{2,-1}}{\partial t} =-i\frac{\Omega}{2} \left(C_{1,0} + C_{1,-2} \right), \\
&\frac{\partial C_{1,0}}{\partial t} = -i\frac{\Omega}{2} \left(C_{2,1} + C_{2,-1} \right),\\ 
&\frac{\partial C_{2,1}}{\partial t} 
= -i\left(2\omega_{21} + \frac{\hbar q^2}{m} \right)C_{2,1} -i \frac{\Omega}{2} \left(C_{1,2} + C_{1,0} \right). 
\end{align}
As long as $\Omega$ is small enough compared to $2\omega_{21} \pm \hbar q^2/m$, which is the case in this experiment,
a Rabi oscillation occurs between the states $|1, k_0\rangle$ and $|2, k_0-q\rangle$ at an angular frequency of $\Omega$.
We thus understand that the case we are now considering is the resonance condition for the transition
between $|1, k_0\rangle$ and $|2, k_0-q\rangle$:
\begin{equation}
\nu_{21}=\frac{\omega_{21}}{2\pi} = \frac{\hbar}{m}\left(k_0-\frac{q}{2} \right)\frac{q}{2\pi} = \frac{\bar{v}}{a},
\end{equation}
\label{RC+}
where
$\bar{v}$ is the atomic velocity averaged before and after the transition.

It is worth mentioning that the resonance condition can be regarded as the 
conservation law of the total energy, the internal energy plus the translational energy, 
of the atom before and after the transition. That is,
\begin{equation}
\frac{\hbar^2k_0^2}{2m} = \hbar \omega_{21}+ \frac{\hbar^2(k_0-q)^2}{2m},
\end{equation}
which indicates that the internal excitation occurs at the expense
of the translational energy. 

Similarly, the resonance condition for the transition between $|1, k_0\rangle$ and $|2, k_0+q\rangle$  is
\begin{equation}
\nu_{21}= -\frac{\hbar}{m}\left(k_0+\frac{q}{2} \right)\frac{q}{2\pi} = -\frac{\bar{v}}{a}.
\label{RC-}
\end{equation}
These resonance conditions Eqs. (22) and (\ref{RC-}) are reduced to Eq.~(\ref{RC}) when $|k_0| \gg q$,
which is fulfilled in this experiment.

\subsection{Time-varying spatially periodic field}
\label{time-variation}
When the periodic field oscillates at a
frequency of $f_m$ in time, the 
interaction Hamiltonian may be represented as
\begin{equation}
V_m=\hbar\Omega(|2\rangle\langle1|+|1\rangle\langle2|)\cos(qx)\cos(2\pi f_mt).
\end{equation}
Following the above derivation, we will obtain that, as long as the rotating wave approximation is valid, the resonance condition for the transition between $|1, k_0\rangle$ and $|2, k_0-q\rangle$ is, 
\begin{equation}
\nu_{21}= \frac{\bar{v}}{a}\pm f_m.
\label{RCfmQ}
\end{equation}
From an energy conservation point of view, the change of the internal energy is compensated for by the change of the translational energy and the absorption or emission of a photon of energy $hf_m$.

\section{Experiment}
\label{experiment}
\begin{figure}
\resizebox{8cm}{!}{%
  \includegraphics{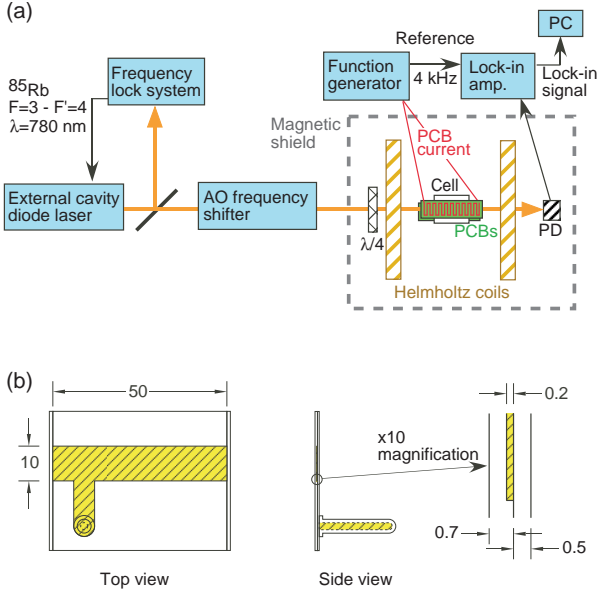}
}
\caption{(a) Schematic diagram of the experimental setup. (b) Drawing of the thin cell. The inside of the cell is indicated by the shaded region. The 10 times magnified image of a part of the cross section (inside the circle, side view) is also shown. The dimensions are in millimeters.}
\label{Fig1}       
\end{figure}

The experimental setup was basically the same as that described in Ref.~\cite{Hat05}.
Figure~\ref{Fig1}(a) shows a schematic diagram of the setup.
An external cavity diode laser provided a laser beam for the $D_2$ transition of
Rb atoms. The laser frequency was locked to the saturation absorption
line of the $F=3 \rightarrow F'=4$ transition of $^{85}$Rb. 
The frequency was then shifted with one or two acousto-optic (AO) modulators by up to 403~MHz,
which corresponded to the Doppler shift of atoms whose velocity component parallel to the laser
beam was 314~m/s.
The laser beam was made circularly polarized with a quarter-wave ($\lambda/4$) plate before entering a thin cell containing
a Rb vapor.
The cell, whose inner dimension was $50 \times 10 \times 0.2$~mm,
was fabricated as shown in Fig.~\ref{Fig1}(b). 
The laser beam was transmitted through the narrow gap of the cell, to which
a periodic magnetic field was applied by sandwiching the cell
with two printed circuit boards (PCBs) having periodic current-carrying traces.
The size of the periodic array was $100 \times 30$~mm, large enough to neglect edge effects.
The produced field essentially had a component normal to the PCBs and its strength varied in a sinusoidal way (period: $a=1.00$~mm) along the laser direction.
The details of the produced field are
explained in Ref.~\cite{Hat05}.
The magnetic resonance within the Zeeman sublevels of the $F=3$ ground
state was detected with a photodiode (PD)
by monitoring variation in the laser absorption
in the cell as scanning a longitudinal magnetic field applied parallel to the laser
beam with a pair of Helmholtz coils.
The Zeeman splitting induced by the longitudinal magnetic field
was 4.67~kHz per $\mu$T.
We employed a lock-in detection scheme by switching on and off 
the current producing the periodic field at 4~kHz.
The cell, PCBs and coils were enclosed
in a magnetic shield. 

\section{Results and discussion}
\label{results}
\begin{figure}
\resizebox{6cm}{!}{%
  \includegraphics{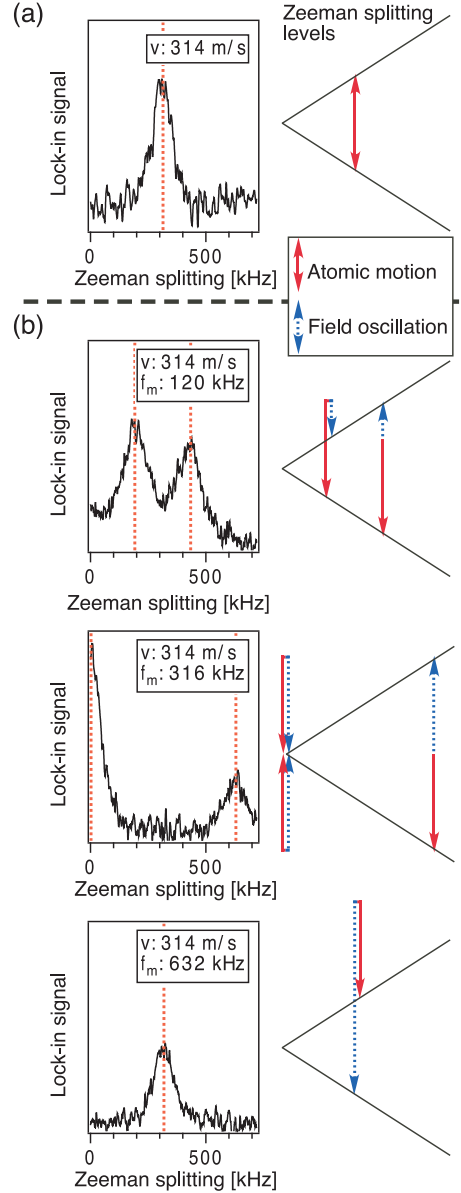}
}
\caption{(a) Resonance spectrum obtained for a laser detuning of 403~MHz and hence a selected velocity of 314~m/s. The resonance condition (Eq.~(\ref{RC})) is indicated by the dotted line. The right diagram schematically shows that the resonance transition is induced by atomic motion between the Zeeman sublevels whose splitting increases with increasing longitudinal magnetic field from left to right. (b) Resonance spectra obtained by oscillating the periodic field at frequency $f_m$ for atoms of velocity 314~m/s. The resonance conditions (Eq.~(\ref{RCfmC})) are indicated by the dotted lines. The signal scales are the same for all the three figures. The right diagrams schematically show that the resonance transitions are induced by the combination of atomic motion and field oscillation. From an energy conservation point of view, the transitions upward (downward) on the red solid arrows represent the decreases (increases) of the translational energy during the transitions, while on the blue dotted arrows indicating the absorptions (emissions) of rf photons.}
\label{Fig2}       
\end{figure}

\begin{figure}
\resizebox{8cm}{!}{%
  \includegraphics{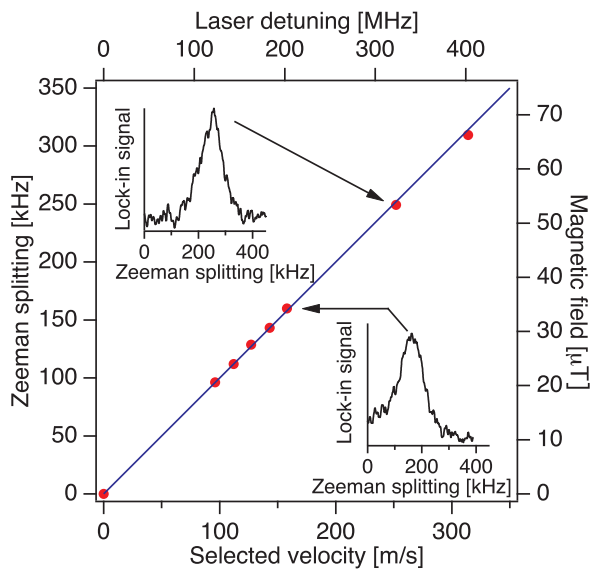}
}
\caption{Zeeman splitting at which the resonance occurs for a static periodic field as a function of selected velocity. The right and top scales show the corresponding longitudinal magnetic field and laser detuning, respectively. The line represents the resonant condition Eq.~(\ref{RC}). Typical resonance spectra are shown in the insets.}
\label{Fig3}       
\end{figure}

We first measured the resonance in a static periodic field. Figure~\ref{Fig2}(a) shows a resonance
spectrum obtained for a laser detuning of 403~MHz, namely, for a
selected velocity of 314~m/s.
The resonance should therefore occur at 314~kHz of Zeeman splitting, 
which was actually observed as seen in Fig.~\ref{Fig2}(a). 
The obtained resonance profile is well fitted by a Lorentzian function.
It is worthwhile to point out that, although the resonant profile is
broadened by the power broadening due to the laser and periodic fields~\cite{Hat05}, the observed linewidth (half width at half maximum: 46~kHz) indicates the coherent interaction of atoms with about three periods of the sinusoidal field, estimated similarly with the transit-time broadening~\cite{Dem96}. This traveling length is 3~mm, much longer than the dimension of the narrow gap of the cell (0.2~mm).
This is the effect of a velocity selection by the thin cell~\cite{Bri99}.

Figure \ref{Fig3} shows the relation between the resonance frequency and the selected velocity. The data were taken as changing the laser detuning. The figure clearly confirms
the resonance condition Eq.~(\ref{RC}) in the investigated range of
0~m/s to 314~m/s.

We then oscillated the PCB current which produced the periodic magnetic field at a frequency of $f_m$ and varied the spatially periodic
field in time as well. Obtained spectra are shown in Fig.~\ref{Fig2}(b) for a selected velocity of 314~m/s.
The single resonance peak obtained for the static field, as shown in Fig.~\ref{Fig2}(a), splits
into two peaks separated
by $2f_m$, as shown in the top graph of Fig.~\ref{Fig2}(b). 
The resonance frequencies are as predicted by Eq.~(\ref{RCfmC}).
Spectra obtained for higher oscillation frequencies are shown
in the middle and bottom graphs of Fig.~\ref{Fig2}(b). 
When $f_m$ nearly equals to the resonance
frequency for the static field, as in the case of the middle graph of Fig.~\ref{Fig2}(b),
the resonance occurs also
near zero Zeeman splitting.
When the oscillation frequency is double the static field resonance frequency,
almost the same spectrum as in the static field is obtained
(the bottom graph of Fig.~\ref{Fig2}(b)).
These behaviors are in good agreement with the resonance condition Eq. (\ref{RCfmC}).
Note that in the middle graph of Fig.~\ref{Fig2}(b) the peak near 0~kHz is much higher than the peak at 630~kHz.
This is mainly because only one of the two counterrotating terms in the interaction Hamiltonian contributes to the resonance at 630~kHz (the rotating wave approximation), while both do near 0~kHz.

Spectroscopy by temporally oscillating the spatially
periodic field may be regarded, on one hand, 
as 
the amplitude modulation spectroscopy of motion-induced resonance in the static
periodic field.
On the other hand, this resonance can be considered a unique rf resonance which is
velocity-selective or Doppler-shifted;
recall that
the ordinary rf resonance spectroscopy in this frequency range does not usually consider the Doppler shift.
Another important point of view is the energy conservation,
which is discussed in the following by taking Fig.~\ref{Fig2}(a) and the bottom graph of
Fig.~\ref{Fig2}(b) as an example:
Although they show similar resonance profiles, the
physical processes of the resonance transition are different.
In the case of Fig.~\ref{Fig2}(a), excitation (de-excitation) in the
internal state occurs by decreasing (increasing)
the translational energy as discussed in Sect.~\ref{quantum}.
In the case of the bottom graph of Fig.~\ref{Fig2}(b), excitation (de-excitation)
occurs by absorbing (emitting) the rf photon energy and further increasing (decreasing)
the translational energy, as discussed in Sect.~\ref{time-variation}.  These relations are schematically shown
in the right diagrams of Fig.~\ref{Fig2}.

These resonance processes may be used for cooling atoms; repeating the cycle of the excitation of the type Fig.~\ref{Fig2}(a)
and the de-excitation of the type Fig.~\ref{Fig2}(b) bottom graph, for example, one can reduce the translational energy of the atom. The proof-of-principle demonstration of this new cooling scheme will be a future subject.

\section{Conclusion}
\label{conclusion}
In this paper we have reported 
resonance transitions between the Zeeman sublevels of optically-polarized Rb atoms traveling through a spatially periodic magnetic field.
The periodic field (period: 1~mm) was applied to a Rb vapor in a thin cell using a pair of printed circuit boards with periodic current-carrying traces printed. We measured resonance spectra as a function of atomic velocity, which was selected up to $\sim300$~m/s using the Doppler effect of the pumping and detecting laser.
The atomic motion induced the resonance when the Zeeman splitting frequency was equal to the atom velocity divided by the field period.
Additional temporal oscillation of the spatially periodic field split a motion-induced resonance peak into two by an amount of the oscillation frequency. 
This resonance can be regarded as velocity-selective rf resonance, particularly at high oscillation frequencies. 
The energy budget of the resonance among the atomic internal energy, translational energy and the photon energy of the oscillating field was discussed.
These experimental results and discussion were supported by
the formulation presented for describing of the resonance induced by atomic motion in a periodic field. Treating the atomic motion 
as a classical trajectory gave an intuitive understanding of the resonance condition, while the quantum mechanical treatment
naturally derived the energy conservation relationship and was particularly important when the change of the translational state was considered.

\begin{acknowledgement}
This work was supported by KAKENHI (Nos. 17684023 and 20684017) and 
Special Coordination Funds for Promoting Science and Technology 
from The Ministry of Education, Culture, Sports, Science and Technology.
\end{acknowledgement}

\appendix
\section{Other formalisms to describe motion-induced resonance in a static periodic field}
This appendix briefly explains two other theoretical treatments of motion-induced resonance.
\subsection{Dressed atom approach}
The quantum mechanical treatment of motion-induced resonance, described in Sect.~\ref{quantum}, can be interpreted in terms of the dressed atom
approach~\cite{Coh92}, which is a powerful method to describe atom--photon interactions.
The standard dressed atom approach considers the global Hamiltonian including the atomic
internal state, photons and atom--photon interaction.
In motion-induced resonance the photon energy term in the Hamiltonian is replaced by the translational energy of the atom. The eigenstates for the system without the interaction are $|j, N\rangle$ in the dressed atom approach,
where $N$ is the number of the photons; for motion-induced resonance
they are $|j, k_0+nq\rangle$.
The increase or decrease of $N$ by one corresponds to the emission or absorption of a photon with an energy of $\hbar \omega_p$ ($\omega_p$: the frequency of photons) by the atom,
while the increase or decrease of $n$ by one is the increase or decrease of the translational momentum of the atom by $\hbar q\bar{v}$. 
One may regard this discrete momentum and energy changes as the emission or absorption of ``quantum'' called Doppleron,
which was introduced to describe the processes regarding the Doppler effect
for atoms in the standing light wave~\cite{Kyr77}.
Furthermore, in both cases, the interaction terms in the Hamiltonians have the matrix elements of the same form
in the basis of these eigenstates.
In this context, motion-induced resonance can be formally treated in the same way as the dressed atom approach.
It is actually very useful to treat a certain type of problem in this way~\cite{Nak08}.

\subsection{General inertial frame of reference}
In this paper we have discussed the resonance processes in the
frame to which the periodic field is fixed.
We off course can adopt a general inertial frame of reference. The new Hamiltonian includes the
source of the periodic field (such as the printed circuit boards in the present
experiment) as
\begin{gather}
H_0 = \hbar \omega_{21}|2\rangle\langle2| +\frac{\hat{p}^2}{2m} +  \frac{\hat{P}^2}{2M}, \notag \\
V =  \hbar\Omega(|2\rangle\langle1|+|1\rangle\langle2|)\cos [q(\hat{x}-\hat{X})],
\end{gather}
where $\hat{P}, M, \hat{X}$ is the translational momentum operator, mass, and the position operator
for the source of the periodic field, respectively. Here the source of the field has been assumed to move
freely.
The eigenstates for $H_0$ now also include the 
momentum of the source of the field.
The following discussion is very similar to that for the field rest frame.
It is interesting to point out that it is easy to show that, in the atom rest frame (strictly speaking, before the transition), the atomic internal excitation
occurs at the expense of the translational energy of the source of the periodic field.

%
%
%
%

\end{document}